\documentclass[twocolumn,showpacs,amsmath,amssymb]{revtex4}
\usepackage{graphicx}
\usepackage{bm}

\begin{document}
 
\title{Thermal analysis of production of resonances in 
relativistic heavy-ion collisions}

\author{Wojciech Broniowski} \email{b4bronio@cyf-kr.edu.pl}
\affiliation{The H. Niewodnicza\'nski Institute of Nuclear Physics,
  PL-31342 Krak\'ow, Poland} \author{Wojciech Florkowski}
\email{florkows@amun.ifj.edu.pl} 
\affiliation{The H. Niewodnicza\'nski Institute of Nuclear Physics,
  PL-31342 Krak\'ow, Poland} 
\affiliation{Institute of
  Physics, \'Swi\c{e}tokrzyska Academy, PL-25406 Kielce, Poland}
\author{Brigitte Hiller}
\email{brigitte@teor.fis.uc.pt} \affiliation{Centro de F\'{\i}sica
  T\'eorica, University of Coimbra, P-3004 516 Coimbra, Portugal}

\vspace{1cm}

\date{10 June 2003}

\begin{abstract}
  Production of resonances is considered in the framework of the
  single-freeze-out model of ultra-relativistic heavy ion collisions.
  The formalism involves the virial expansion, where the probability
  to form a resonance in a two-body channel is proportional to the
  derivative of the phase-shift with respect to the invariant mass.
  The thermal model incorporates longitudinal and transverse flow, as well
  as kinematic cuts of the STAR experiment at RHIC.  We
  find that the shape of the $\pi^+ \pi^-$ spectral line qualitatively
 reproduces the preliminary experimental data when the position of the 
$\rho$ peak is lowered. This confirms the need to include the medium 
effects in the description of the RHIC data. We also analyze the
  transverse-momentum spectra of $\rho$, $K^*(892)^0$, and
  $f_0(980)$, and find that the slopes agree with the observed
  values. Predictions are made for $\eta$, $\eta^\prime$, $\omega$, $\phi$,
  $\Lambda(1520)$, and $\Sigma(1385)$.

\end{abstract}



\maketitle 

\section{Introduction \label{intro}}

With the help of the mixed-event technique the STAR Collaboration has
managed to obtain preliminary results on the production of hadronic
resonances.  The data were presented for $K^*(892)^0$
\cite{starKstar1,starKstar2}, $\rho(770)^0$
\cite{patricia1,patricia2}, $f_0(980)$ \cite{patricia1,patricia2}, and
$\Lambda(1520)$ \cite{christelle,markert,ludovic}.  The invariant-mass
distribution of the particles produced in the decays of the resonances
may be used to get the information on the possible in-medium
modifications of hadron masses. Such modifications are predicted by
various theoretical models of dense and hot hadronic matter
\cite{BR,hatlee}, as well as hinted by the enhancement of the dilepton
production in the low-mass region \cite{CERES,HELIOS}.  Most
interestingly, the measured invariant-mass distribution of the
$\pi^+\pi^-$ pairs \cite{patricia1,patricia2} also suggests a drop of
the effective mass of the $\rho$ meson by several tens of MeV
\cite{patricia1}.  This effect was recently discussed 
by Shuryak and Brown \cite{shuryakbrown}, Kolb and Prakash \cite{KolbP}, 
and Rapp \cite{Rapp}, with the conclusion drawn that it is a
genuine dynamical effect induced by the interaction of the $\rho$ meson
with the hadronic matter.  On the other hand, the measurement of the
invariant-mass distributions of the $\pi K$ pairs indicates that the
effective mass of the $K^*(892)^0$ remains unchanged
\cite{starKstar2}.

Since one measures the properties of stable hadrons which move freely
to detectors, the experimentally obtained correlations may bring us information only
about the final stages of the evolution of the hadronic matter, {\em
  i.e.}, about the conditions at freeze-out. From this point of view
it is interesting to analyze the production of resonances in the
framework of the thermal approach (the single-freeze-out model of
Refs.  \cite{wbwf,str,zakop}) which turned out to be very successful
in reproducing the yields and spectra of stable hadrons.
In particular, the thermal model can be naturally used to study the
impact of the possible in-medium modification of the $\rho$-meson mass on
different physical observables, {\em e.g.}, on the correlation in the
invariant mass of the $\pi^+\pi^-$ pairs, the ratios of the resonance
abundances, or their transverse-momentum spectra.

The paper is organized as follows: In the next Section we outline the
Dashen-Ma-Bernstein formalism used to describe a gas of hadronic
resonances. In Sec. \ref{statsource} we present the invariant-mass
correlations of $\pi^+\pi^-$ pairs emitted from a static thermal
source, and in Sec. \ref{chemkin} we discuss the effect of the
temperature of such a source on the shape of the spectral $\pi^+\pi^-$
line. In Sec. \ref{sec:flow} we present the main assumptions of the
single-freeze-out model. With the knowledge of both the experimental
kinematic cuts (Sec. \ref{kincuts}) and the feeding from the decays of
higher resonances (Sec. \ref{higher}), the model is then used to
compute the invariant mass spectrum of the $\pi^+\pi^-$ pairs (Sec.
\ref{results}). In Sec. \ref{sec:ratios} and \ref{sec:trspectra} we
present the model results for the ratios of the resonance yields and
for the resonance transverse-momentum spectra (wherever it is possible
we compare the model results with the preliminary data). The paper
contains three Appendices which give the parameters for the
$\pi^+\pi^-$ phase shifts and explain in simple terms the implementation of the
experimental kinematic cuts in our calculations of two- and three-body
decays.

\section{Production of resonances and the phase-shifts}

The formalism for the treatment of resonances in a hadronic gas in
thermal equilibrium has been developed by Dashen, Ma, and Bernstein
\cite{Da1}, and Dashen and Rajaraman \cite{Da2}, and in the context of
heavy-ion physics has been recalled and further elaborated by
Weinhold, Friman, and N\"orenberg
\cite{Wmsc,WFNapp,WFNnote,WFNplb,Wphd} (see also
\cite{denis,larionov,pelaez}). The basic formula following from the
formalism is that density of the resonance per volume and per unit
invariant mass, $M$, produced in the two-body channel of particles 1
and 2 in thermal equilibrium is given by the formula
\begin{eqnarray}
\frac{dn}{dM} = f \int \frac{d^3 p}{(2\pi)^3} \frac{1}{\pi} 
\frac{d\delta_{12}(M)}{dM} 
\frac{1}{\exp \left( \frac{\sqrt{M^2+{\bm p}^2}}{T} \right ) \pm 1}, 
\label{master}
\end{eqnarray} 
where $\delta_{12}(M)$ is the phase shift for the scattering of
particles 1 and 2, $f$ is a spin-isospin factor, $T$ is the
temperature, and the sign in the distribution function depends on the
statistics. In practice, one may replace the distribution function by
the Boltzmann factor, since the effects of the quantum statistics are
numerically small.

As pointed out by the authors of Ref.~\cite{WFNnote}, in many works
the spectral function of the resonance is used {\em ad hoc} as the
weight in Eq.~(\ref{master}) instead of the derivative of the phase
shift, which is the correct thing to do. 
For narrow resonances this does not make a difference, since
then both the spectral function and the derivative of the phase shift
peak very sharply at the resonance position, $m_R$, {\em i.e.},
$d\delta_{12}(M)/dM \simeq \pi \delta(M-m_R)$, and then one recovers
the narrow-resonance limit
\begin{eqnarray}
n^{\rm (narrow)} = f \int \frac{d^3 p}{(2\pi)^3} 
\frac{1}{\exp \left( \frac{\sqrt{m_R^2+{\bm p}^2}}{T} \right ) \pm 1}. 
\label{narrow}
\end{eqnarray} 
However, for wide resonances, or for effects of tails, the difference
between the correct formula (\ref{master}) and the one with the
spectral function is very significant, not only conceptually but also
numerically.

We first focus on the $\pi^+\pi^-$ scattering. We use the
experimental phase shifts, which can be parameterized with simple
formulas of Ref.~\cite{pipipar}:
\begin{eqnarray}
&&\tan \left ( \delta^I_l(M) \right )=\sqrt{1-m_\pi^2/M^2} \; q^{2l} 
\label{tandel}\\
&& \times \frac{(A^I_l+B^I_l q^2+C^I_l q^4+D^I_l q^6)(4m_\pi^2-s^I_l)}
{M^2-s^I_l}, \nonumber
\end{eqnarray}
where $q=\sqrt{M^2-4m_\pi^2}/2$, $m_\pi=139.57$~MeV is the mass of the
charged pion, and the remaining parameters are listed in App.
\ref{app:param}. The relevant channels are $I=1$, $l=1$ ($\rho$),
$I=0$, $l=0$ ($f_0$/$\sigma$), and $I=2$, $l=0$.

\begin{figure}[b]
\begin{center}
\includegraphics[width=8.3cm]{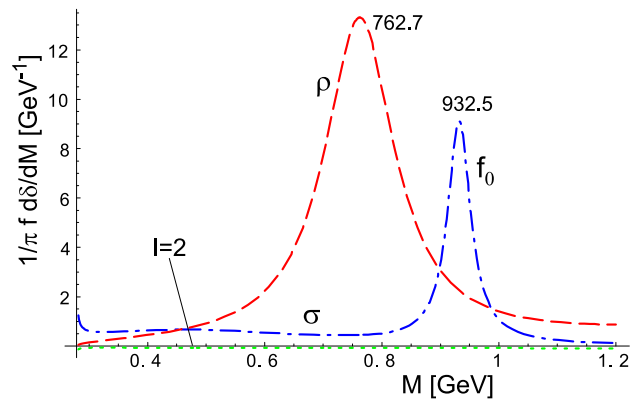}
\end{center}
\caption{The weight functions, $1/\pi f\,d\delta/dM$, in the isoscalar, isovector, 
  and isotensor channels, plotted as a function of the $\pi^+ \pi^-$
  invariant mass.  The phase shifts are taken from the
  parameterization of the experimental data \cite{pipipar}.  Numerical
  labels indicate the peak positions (in MeV) which correspond to
  inflection points in the phase shifts as functions of $M$. The $I=2$
  channel is negligible.}
\label{phase}
\end{figure}

Figure \ref{phase} shows the weights in Eq.~(\ref{master}), {\em
  i.e.}, the quantities $f d\delta(M)/(\pi dM)$, plotted as functions
of $M$.  The factor $f$ consists of the spin degeneracy, $2l+1$, and
the isospin Clebsch-Gordan coefficient, equal $2/3$ for the
isosinglet, $1$ for the isovector, and $1/2$ for the $I=2$ channel. In
the isovector channel the peak of the weight function is found at
$762.7$~MeV, which corresponds to the inflection point in $\delta(M)$,
and not the point where $\delta(M)=\pi/2$, which is at $773.6$~MeV
(note a 10~MeV ``dropping'' of the $\rho$-meson ``mass''). For the
isoscalar channel the corresponding curve in Fig.~\ref{phase} peaks at
932.5~MeV. We note that in this channel the weight function $f d\delta/(\pi dM)$ is
significantly different from the spectral function. The latter one appears
in the calculation of the production {\em rate} (see for instance
\cite{KolbP,Rapp}), whereas our study concerns the invariant-mass spectra,
where formula (1) holds. As a result of the use of the phase-shifts, 
there is a very small contribution from the $\sigma$ meson,
since the strength of the scalar channel, as seen in Fig.~\ref{phase},
is small. It is worthwhile to notice that it peaks at the two-pion
threshold \cite{pipipar}, which is an immediate consequence of Eq.
(\ref{tandel}) for $l=0$. The contribution of the $I=2$ channel is tiny
(also note that it is negative since the phase shift in this channel
is a decreasing function of $M$, {\em cf.} \cite{Da2}).

For three-body decays formulas analogous to Eq.~(\ref{master}) may be
given, {\em cf.}  Ref.~\cite{Da1,Da2}. They would involve the detailed
dynamical information on the decay process. For our purpose this is
not necessary (the three-body decays feed the range of rather low
invariant $\pi^+ \pi^-$ masses) and also not practical, since the
detailed information on the dynamical dependences of the appropriate
transition matrices is not easy to extract from the experimental data.
In our calculations we include the resonances decaying into three-body
channels which are very narrow ($\omega$, $\eta$, and $\eta'$). Thus,
as is customary in similar applications, we only account for the
phase-space dependence on the invariant mass $M$, and do not consider
dynamical effects of the transition matrix.
 
\section{Resonances in the static source \label{statsource}}

First, in order to gain some experience, we consider the simplest case
of the emission of particles from a static source. We will come back
to a more realistic description incorporating the flow in
Sec.~\ref{sec:flow}. We also assume in this Section the
single-freeze-out hypothesis \cite{wbwf,str,zakop}, hence no effects
of rescattering are incorporated after the chemical freeze-out.  We
thus assume the chemical freeze-out temperature to be
\cite{wfwbmm,BM4,review}
\begin{equation}
T_{\rm chem}=165~{\rm MeV}, \label{Tchem}
\end{equation}    
and, as we have said, the thermal freeze-out occurs at the same
temperature, $T_{\rm therm}=T_{\rm chem}$. In the next Section we will
lift this assumption.

We now simply use Eq.~(\ref{master}) with $T=T_{\rm chem}$, and
include the following resonances that couple to the $\pi^+ \pi^-$
channel: $\rho$, $f_0$/$\sigma$, $\omega$, $\eta$, $\eta^\prime$,
$K_S$, and $f_2(1270)$. For the omega, both the three-body mode,
$\omega \to \pi^+ \pi^- \pi^0$, and the two-body mode, $\omega \to
\pi^+ \pi^-$, are incorporated. The appropriate branching ratios are
included in the constant $f$ in Eq.~(\ref{master}).  For channels
other than those of Fig.~\ref{phase}, where the experimental phase
shifts are included, we use the simple Breit-Wigner parameterization,
which is good for the narrow resonances. For the width of $K_S$ we
take the experimental resolution of the STAR experiment, which is about
$10$ MeV \cite{FachPriv}. The width of the $\omega$ is also increased
by the same value.

We compute the spectra at mid-rapidity, hence we use
\begin{eqnarray}
\left .\frac{dn}{dMdy} \right |_{y=0} &=& 
\sum_i f_i \int_{p_\perp^{\rm low}}^{p_\perp^{\rm high}} 
\frac{p_\perp dp_\perp}{(2\pi)^2} \frac{1}{\pi} \frac{d\delta_{i}(M)}{dM} 
\label{used} \\
&& \times \frac{\sqrt{M^2+p_\perp^2}}
{\exp \left( \frac{\sqrt{M^2+{\bm p_\perp}^2}}{T} \right ) - 1}, \nonumber
\end{eqnarray} 
where $i$ labels the channel, $n$ is the volume density of the $\pi^+
\pi^-$ pairs, and $M$ is the invariant mass of the $\pi^+ \pi^-$
system.  The above formula, written for two-body decays, is
supplemented in the actual calculation with the three-body reactions.
The limits for the transverse-momentum integration are taken at the
upper and lower cuts for the STAR experiment \cite{patricia2},
$p_\perp^{\rm low}=0.2~{\rm GeV}$ and $p_\perp^{\rm high}=2.2~{\rm
  GeV}$.

\begin{figure}[b]
\begin{center}
\includegraphics[width=8.7cm]{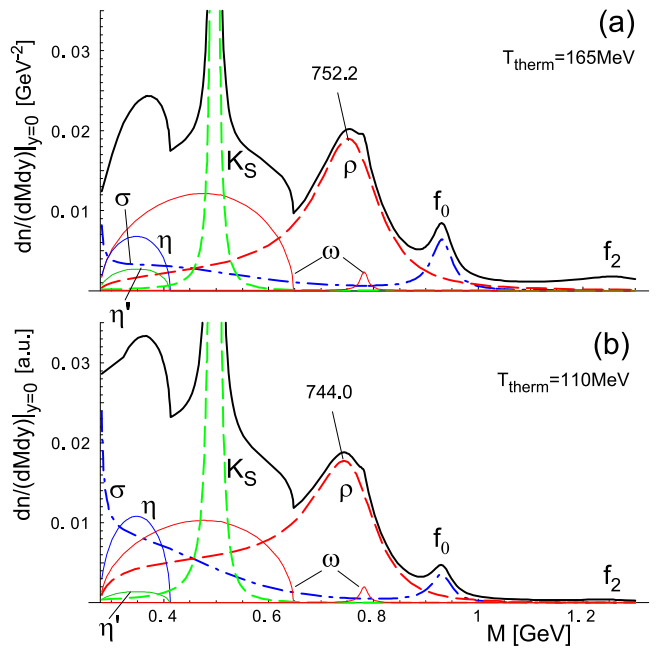}
\end{center}
\caption{The volume density of the $\pi^+ \pi^-$ pairs 
  plotted as a function of the invariant mass, obtained from the
  thermal model with a static source, Eq.~(\ref{used}).  (a) The case
  of $T_{\rm therm}=T_{\rm chem}=165~{\rm MeV}$.  (b) The case of
  $T_{\rm therm}=110~{\rm MeV}$. The numbers indicate the position of
  the $\rho$ peak (in MeV). The labels indicate the decays included,
  with the two-body $\pi^+ \pi^-$ channels and the three-body
  channels, $\omega \to \pi^0 \pi^+ \pi^-$, $\eta \to \pi^0 \pi^+
  \pi^-$, and $\eta' \to \eta \pi^+ \pi^-$. The relative contributions
  from the resonances are fixed by the value of the temperature.}
\label{static}
\end{figure}

The results of the calculation are shown in Fig.~\ref{static}(a). The
two-body resonances lead to well-visible peaks, while the three-body
channels of the $\omega$, $\eta$ and $\eta'$ produce typical broad
structures at lower values of $M$. Note that the scalar-isoscalar
channel generates the resonance, $f_0$, and a smooth background, $\sigma$.
This background increases with the decreasing $M$, and peaks at the
threshold.

\section{Chemical vs. kinetic freeze-out \label{chemkin}}

In this section we analyze in simple terms the effects of possible
lower temperature of the thermal freeze-out on the shape of the $\pi^+
\pi^-$ invariant-mass correlations. To this end we simply use
Eq.~(\ref{used}) at a lower value of $T$.

Consider, for brevity of notation, the gas consisting only of $\pi^+$,
$\pi^-$, and $\rho^0$.  We assume isospin symmetry for the system,
which works very well for RHIC \cite{brahmsy}. The pions rescatter
elastically through the $\rho$-channel. The total number of pions
that are observed in the detectors is equal to
\begin{eqnarray}
N_\pi &=& V^{\rm chem}\left ( 2 \times n_\pi(T_{\rm chem},\mu=0) \right . 
\nonumber \\
&+& \left . 2 \times 3 \times n_\rho(T_{\rm chem},\mu=0) \right ), 
\label{pichem}
\end{eqnarray}
where 
\begin{eqnarray}
n_i(T,\mu)=\int \frac{d^3 p}{(2\pi)^3}
\exp \left( \frac{-\sqrt{m_i^2+{\bm p}^2}+\mu}{T} \right ). 
\end{eqnarray}
The factor of 2 in front of $n_\pi$ in Eq.~(\ref{pichem}) comes from
the $\pi^+$-$\pi^-$ degeneracy, the factor of 2 in front of $n_\rho$
comes from the fact that rho decays into two pions, and the factor of
3 is the spin degeneracy of the $\rho$.  The chemical potential,
$\mu$, vanishes at the chemical freeze-out. The volume at the chemical
freeze-out is denoted by $V^{\rm chem}$.

The chemical freeze-out is defined as the stage  in the evolution
when the abundances of {\em stable} (with respect to the strong
interaction) particles have been fixed. Next, the system expands, and
while it progresses in its evolution, the elastic scattering may
occur. This scattering proceeds through the formation of unstable
resonances, until the system is too dilute for the scattering to be
effective and the thermal freeze-out occurs. At the moment of the
thermal freeze-out the total number of pions in our example is
\begin{eqnarray}
N_\pi &=& V^{\rm therm}\left ( 2 \times n_\pi(T_{\rm therm},\mu_\pi) \right . 
\nonumber \\
&+& \left . 2 \times 3 \times n_\rho(T_{\rm therm},\mu_\rho) \right ), 
\label{pitherm}
\end{eqnarray}
where $V^{\rm therm}$ is the volume at the thermal freeze-out,
$\mu_\pi$ and $\mu_\rho=2 \mu_\pi$ (the reaction $\rho^0
\leftrightarrow \pi^+ \pi^-$ is in equilibrium) are the chemical
potentials which ensure that $N_\pi$ is conserved, as requested by the
fact that the system had frozen chemically. Knowing the ratio $V^{\rm
  therm}/V^{\rm chem}$ we could compute $\mu_\pi$ comparing the
right-hand sides of Eqs.~(\ref{pichem}) and (\ref{pitherm}). However,
for the present purpose, where we are only interested in the shape of
the correlation function in $M$ and not the absolute values, this is
not necessary.  The chemical potential enters in Eq.~(\ref{pitherm})
as a multiplicative constant, $V^{\rm therm} \exp (2\mu_\pi/T_{\rm
  therm})$. Since we do not control in the present calculation the
volume, we can remove the normalization constant from our
consideration. The argument holds if more reaction channels are
present.

Thus, we redo the calculation of the previous Section, based on
Eq.~(\ref{used}), but now with a lower temperature and an arbitrary
normalization constant. The results for $T_{\rm therm}=110$~MeV are
shown in Fig.~\ref{static}~(b). We note a prominent difference from
the case of Fig.~\ref{static}~(a), where the temperature was
significantly higher: the high-$M$ spectrum is suppressed, while the
low-$M$ spectrum is enhanced. Beginning from the high-mass end, we
notice that the $f_2$ resonance has practically disappeared, the
relative strength of the $f_0$ to the $\rho$ peak is only about 1/5
compared to 1/3 in Fig.~\ref{static}~(a), finally, left to the $K_s$
peak we note a significantly larger background from the
$\sigma$ tail, the shoulder of the $\rho$, and the $\eta$ decays.
While for $T=165$~MeV the hight of this background is only slightly
above the hight of the $\rho$ peak, for $T=110$~MeV it rises to
about two times the hight of the $\rho$ peak.

We also remark that due to the presence of the thermal function in
Eq.~(\ref{used}), the position of the peak is shifted downwards from
the original vacuum value to 752.2~MeV for $T_{\rm therm}=165$~MeV,
and to 744.0~MeV for $T_{\rm therm}=110$~MeV.

To conclude this Section, we state that the very simple thermal
analysis shows that the shape of the ``spectral line'' of the $\pi^+
\pi^-$ system depends strongly on the temperature of the thermal
freeze-out, which, when compared to accurate data, may be used to
determine $T_{\rm therm}$ in an independent manner.  In the next
sections we elaborate on this observation by incorporating other
important effects.

\section{Flow and the single-freeze-out model \label{sec:flow}}

The medium produced at mid-rapidity in ultra-relativistic heavy-ion
collisions undergoes a rapid expansion, characterized by the
longitudinal and transverse flow.  Although flow has no effect on the
invariant mass of a pair of particles produced in a resonance decay,
since the quantity is Lorentz-invariant, it nevertheless affects the
results, since the kinematic cuts imposed in the experiment in an
obvious manner break this invariance.  In Ref.~\cite{wbwf,str} we have
constructed a thermal model which includes the flow effects. The model
has the following main ingredients:
\begin{enumerate}
\item There is one universal freeze-out, occurring at the temperature
\begin{equation}
T_{\rm chem}=T_{\rm therm} \equiv T \simeq 165~{\rm MeV}. \label{Tuniv}
\end{equation}
In \cite{wfwbmm,wbwf,str,zakop} we have demonstrated that with the
inclusion of all hadronic resonances the distinction between the 
traditionally considered chemical and thermal freeze-outs
is not necessary, at least for RHIC.

\item To describe the geometry and flow at the freeze-out we adopt the
  approach of
  Refs.~\cite{bjorken,baym,Kolya,siemens,SSH,BL,cs1,Rischke,SH,cs2}.
  The freeze-out hypersurface is defined by the condition
\begin{equation}
\tau = \sqrt{t^2-r^2_x-r^2_y-r^2_z} = {\rm const},
\label{tau}
\end{equation}
while the transverse size of the fire-cylinder is made finite 
by requesting that
\begin{equation}
\rho \equiv \sqrt{r_x^2+r_y^2}< \rho_{\rm max}.
\label{rhodef}
\end{equation}
In addition, we assume that the four-velocity of the 
expansion at freeze-out is proportional to the coordinate, namely
\begin{equation}
u^{\mu } =\frac{x^{\mu }}{\tau }=\frac{t}{\tau }\left(
1,\frac{r_{x}}{t},\frac{r_{y}}{t},\frac{r_{z}}{t}\right).
\label{umu}
\end{equation}
The model is explicitly boost-invariant, which in view of the recent
data delivered by BRAHMS \cite{brahmsy} is justified for the
description of particle production in the rapidity range $-1<y<1$.

\item All resonances from the Particle Data Tables \cite{PDG} are
  included in the calculation, which is important due to the
  Hagedorn-like behavior of the resonance mass spectrum
  \cite{hagedorn,myhag,bled,rafhag}.

\end{enumerate}

The model has altogether four parameters: two thermal parameters, $T$,
and the baryon chemical potential, $\mu_B$, which are fitted to the
available particle ratios, and two geometric parameters, $\tau$ and
$\rho_{\rm max}$, fitted to the transverse-momentum spectra.  The
details of the model can be found in Ref.~\cite{zakop}. The model
works very well and economically for the particle ratios
\cite{wfwbmm}, transverse-momentum spectra \cite{wbwf}, as well as
strange-particle production \cite{str}.

\begin{figure*}[tb]
\begin{center}
\includegraphics[width=17.5cm]{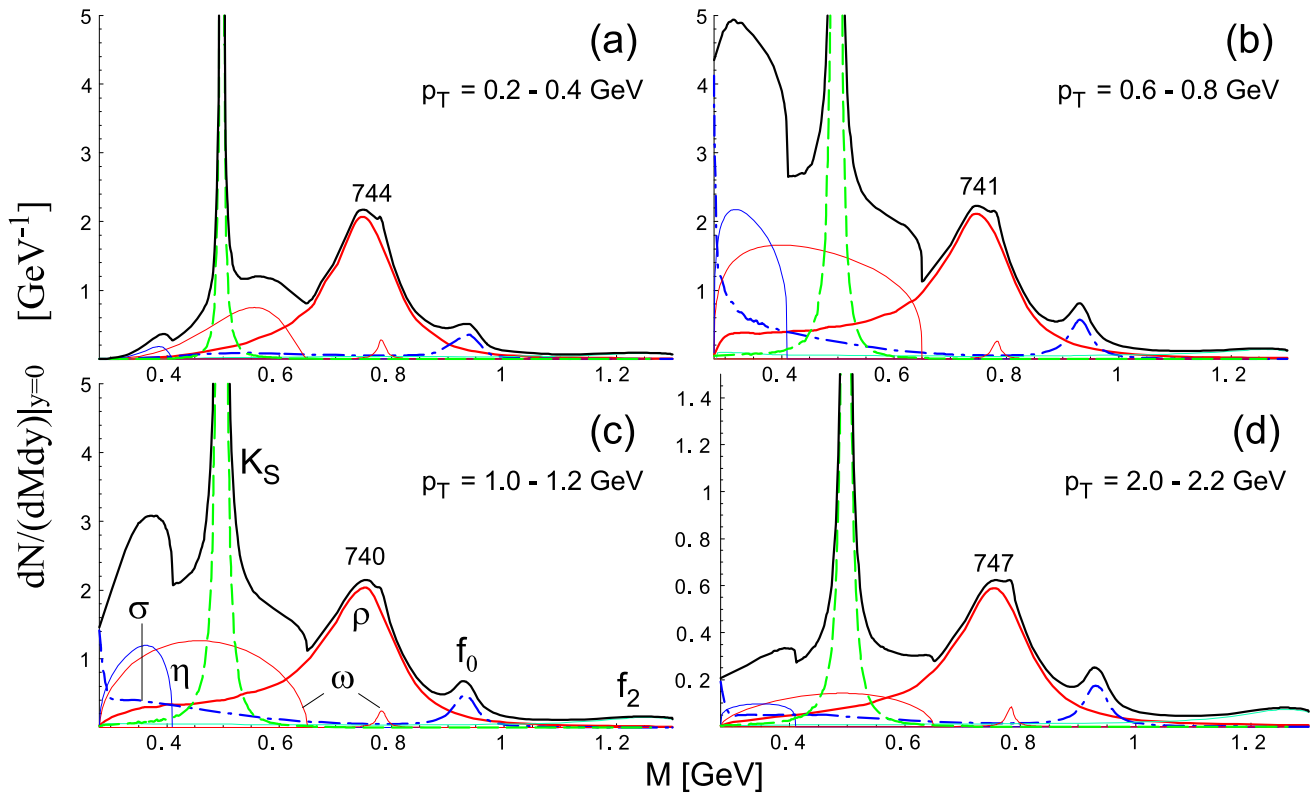}
\end{center}
\caption{The invariant $\pi^+ \pi^-$ mass spectra in the single-freeze-out model 
  for four sample bins in the transverse momentum of the pair, $p_T$,
  plotted as a function of $M$.  The contribution of various
  resonances are labeled in Fig.~(c), with $\eta$ indicating the joint
  contribution of $\eta$ and $\eta'$. The kinematic cuts of the STAR
  experiment are incorporated. The labels at the $\rho$ peak indicate
  its position in MeV, lowered as compared to the vacuum value due to
  thermal effects.}
\label{fig:sfm}
\end{figure*}

\section{Kinematic cuts \label{kincuts}}

The next important effect is related to the experimental cuts and
traditionally has been the domain of experimentalists. However, in the
present application the inclusion of all relevant kinematic cuts of
the STAR analysis \cite{patricia1,patricia2,FachPriv} can be included
straightforwardly.  Needless to say that the proper inclusion of
kinematic constraints is frequently crucial when comparing theoretical
models to the data.

For two-body decays, the relevant formula for the number of pairs of
particles $1$ and $2$, derived in App. \ref{app:two}, has the form
\begin{eqnarray}
\frac{dN_{12}}{dM} &=& \frac{d\delta _{12}}{dM} 
{\frac{b m}{p_{1}^{\ast }}}
\int_{p_{1, \rm{low}}^{\perp }}^{p_{1, {\rm high}}^{\perp }}
\!\!\!\!\! dp_{1}^{\perp }\int_{y_{1, {\rm low}}}^{y_{1, {\rm high}}} 
\!\!\!\!\! dy_{1}\int_{p_{\rm low}^{\perp }}^{p_{\rm high}^{\perp }} 
\!\!\!\!\! dp^{\perp } \int_{y_{\rm low}}^{y_{\rm high}} 
\!\!\!\!\! dy  \nonumber \\
&&\times C_{2}^{0} C^\eta_1 C^\eta_2 
\,\frac{\theta (1-\cos^2 \gamma _{0})}{|\sin \gamma _{0}|} S(p^{\perp }),
\label{gener}
\end{eqnarray}
with all quantities defined in App.~\ref{app:two}.

For the case of three-body decays we have
\begin{eqnarray}
\frac{dN_{12}}{dM} &=& b \int_0^\infty 2\pi p^\perp dp^\perp 
\int_{-\infty}^\infty dy  \frac{dN_{12}({\bm p}^\perp, y)}{dM} S(p^{\perp }),
\nonumber \\ ~ \label{3b}
\end{eqnarray}
where $b$ is the branching ratio and $dN_{12}({\bm p}^\perp, y)/dM$
has been evaluated in App.~\ref{app:three}, with full inclusion of all
cuts relevant in the STAR experiment.

The cuts in the STAR analysis of the $\pi^+ \pi^-$ invariant-mass spectra
have the following form \cite{patricia1,patricia2,FachPriv}:
\begin{eqnarray}
 |y_\pi| &\leq& 1, \nonumber \\
 |\eta_\pi| &\leq& 0.8, \label{starc}\\
 p^\perp_\pi &\geq& 0.2~{\rm GeV} , \nonumber
\end{eqnarray}
while the bins in $p_T \equiv |{\bm p}_\pi^\perp+{\bm p}_\pi^\perp |$
start from the range $0.2-0.4$~GeV, and step up by $0.2$~GeV until the
range $2.2-2.4$~GeV.  These conditions are implemented in the
calculation shown below.

\section{Decays of higher resonances \label{higher}}

An important ingredient and, in fact, the key to the success of the
thermal models in both reproducing the particle ratios and the
transverse-momentum spectra \cite{wfwbmm,wbwf,str,zakop}, is the
inclusion of resonances.  Although the thermal distribution suppresses
the high-mass particles, their abundance grows exponentially according to the
Hagedorn hypothesis \cite{hagedorn,myhag,bled,rafhag}, and in practice one
needs to go very high up in the mass of the resonances in order to
obtain stable results \cite{mm}.  We include all resonances from the
Particle Data Tables \cite{PDG}.  The high-lying resonances decay in
cascades, eventually producing stable particles.

The resonances considered in this paper, in particular
the $\rho$, also acquire substantial contributions from the higher
resonances, {\em e.g.}, $\eta' \to \rho^0 \gamma$, or $\phi(1020) \to
\rho \pi$. Such effects, entering at the level of a few tens of \%,
are difficult to account for accurately in our formalism. This is due
to the dynamics characterized by the parameters that are not well known. For
instance, the decay of $\eta'$ may proceed through the $\rho^0
\gamma$ channel, as well as directly into the uncorrelated three-pion
state, $\pi^+ \pi^- \pi^0$, which forms a smooth background around the
$\rho$ peak. In other words, the feeding of the $\rho$ from the higher
resonance need not reproduce the shape of the $\rho$ peak from Fig.~1,
and the resulting dependence on $M$ may be altered to some extent.
This important but, due to experimental uncertainties, not easy issue
is left for later studies. At the moment we take the simplest way, and
assume that the shape of the spectral line in Fig.~1 is not altered. This
amounts to including a multiplicative factor, $d$, for each considered
resonance. These factors are obtained from the thermal model as 
discussed in Ref.~\cite{wfwbmm,zakop}.
The thermal parameters for the full RHIC energy of
$\sqrt{s_{NN}}=200$~GeV are \cite{abwbwf} 
\begin{eqnarray}
T&=&165.6{\rm ~MeV}, \nonumber \\
\mu_B&=&28.5{\rm ~MeV}, \nonumber \\
\mu_S&=&6.9{\rm ~MeV}, \nonumber \\
\mu_I&=&-0.9{\rm ~MeV}.
\end{eqnarray} 
The calculation leads to the following enhancement factors coming from the
decays of higher resonances: $d_{K_S}=1.98$, $d_\eta=1.74$,
$d_\sigma=1.13$, $d_\rho=1.42$, $d_\omega=1.43$, $d_{\eta'}=1.08$,
$d_{f_0}=1.01$, and $d_{f_2}=1.28$.  Thus, the effects is strongest
for light particles, $K_S$, $\eta$, $\rho$, and $\omega$, while it is weaker for 
the heavier $\eta'$ and scalar mesons.

\section{Results of the single-freeze-out model \label{results}}

We may finally come to presenting the results of the calculation in
the framework of the single-freeze-out model.  The calculation
includes the kinematic cuts described in App.~\ref{app:two} and
\ref{app:three}, and the enhancement factors from the higher
resonances, $d_i$, described in Sec.~\ref{higher}. 
The expansion parameters are taken to be $\tau=5$~fm and
$\rho_{\rm max}=4.2$~fm, which according to the fits 
of the $p_\perp$ spectra, corresponds to the centrality 40-80\%
\cite{abwbwf}. 
In Fig.~\ref{fig:sfm} we show the results obtained with the help of
Eqs.~(\ref{gener},\ref{3b}) with the STAR kinematic cuts
(\ref{starc}), for four sample bins in the transverse momentum of the
pair, $p_T$, with the lowest $p_T$ in Fig.~3(a) and highest in
Fig.~3(d).  The contributions of various resonances are clearly
visible. We note that the shape of the spectrum changes with the
assumed $p_T$ bin. For both the lowest and the highest $p_T$ the
low-$M$ region is suppressed, while at intermediate $p_T$ the spectrum
has large contributions at low $M$. In our calculation the relative
contributions form various resonances is {\em fixed}, as given by the
model. Also, the relative normalization between Figs.~(a-d) is
preserved. The labels at the $\rho$ peak indicate its position in MeV.
We observe that it is lowered compared to the vacuum value, due to the
thermal effects, and assumes values between 740 and 747~MeV. This fall
is not as large as that observed in the preliminary STAR data
\cite{patricia1,patricia2}, where the position of the peak resulting
from fitting the data is much lower, about 700~MeV. Our
single-freeze-out model with the vacuum $\rho$ meson is not capable of
producing this result.

This is perhaps the most important outcome of our analysis: both the
naive thermal approach, {\em cf.} Fig.~2, and the full-fledged
single-freeze-out model with expansion and kinematic cuts, are {\em
  unable to reproduce the (preliminary) STAR data if the vacuum value
  of the $I=1$ $\pi \pi$ phase shift (\ref{tandel}) is used.} This
confirms the earlier conclusions of Shuryak and Brown
\cite{shuryakbrown} and Rapp \cite{Rapp}. The model with the vacuum
$\rho$ can bring the peak down to $\sim 740$~MeV, calling for about
additional 40~MeV from other effects, such as the medium
modifications.

\begin{figure}[tb]
\begin{center}
\includegraphics[width=8.7cm]{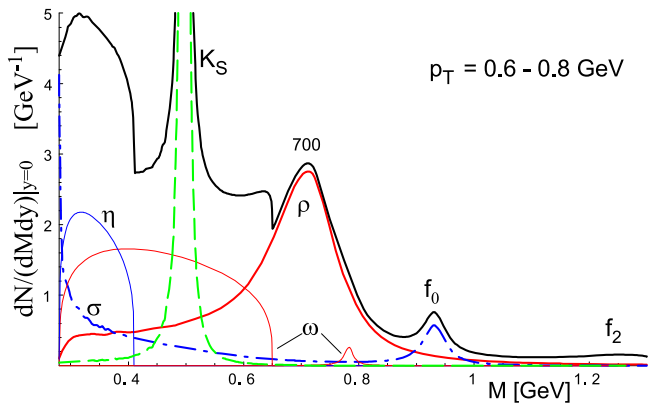}
\end{center}
\caption{Same as Fig.~\ref{fig:sfm}(c) with the $\rho$ meson scaled down by 
  9\%. Now the peak of the $\rho$ has moved to the preliminary value
  observed in the STAR experiment \cite{patricia1,patricia2}. }
\label{fig:sfmscaled}
\end{figure}

In order to show how the medium modifications will show up in the
$\pi^+ \pi^-$ spectrum, we have scaled the $\pi \pi$ phase shift in
the $\rho$ channel, according to the simple law
\begin{eqnarray}
\delta^1_1(M)_{\rm scaled}=\delta^1_1(s^{-1} M)_{\rm vacuum}, \label{sdel}
\end{eqnarray}
where $s$ is the scale factor. We use $s=0.91$, which places the peak
at 700~MeV. The result of this calculation is shown in
Fig.~\ref{fig:sfmscaled}. Now, with the lowered $\rho$ position, the
calculation looks very similar to the preliminary data of
Ref.~\cite{patricia1,patricia2}. With the $\rho$ peak moved to the
left the dip between the $\rho$ and $K_S$ peaks is largely reduced, as
seen in the preliminary data. Also, the background on the left side of the $K_S$
peak, coming from other resonances, is, for the considered $p_T$ bin,
significantly higher than the hight of the $\rho$ peak.

We also remark that the shape of the the curves obtained in the
single-freeze-out model for intermediate values of $p_T$ is closer to
the naive thermal-model calculation of Fig.~2 with $T_{\rm
  therm}=110$~MeV rather than $T_{\rm therm}=165$~MeV. This is
reminiscent of the ``cooling'' effect of the resonance decays, {\em cf.} Ref.~\cite{wfwbmm}.
Indeed, the enhancement factors $d_i$, higher for low mass $M$ and lower 
for high mass $M$, connected with the feeding
of the resonances by even higher excited states, lead to the
modification of the spectral line such that it resembles a lower temperature
spectrum obtained without the resonance feeding.

\begin{table*}[t]
\begin{centering}
\begin{tabular}{|l|c|c|c|}
\hline
&
\begin{tabular}{c} 
Model \\
$m^*_\rho = 770$ MeV \\
\end{tabular} 
& 
\begin{tabular}{c} 
Model \\
$m^*_\rho = 700$ MeV \\
\end{tabular} 
& Experiment \\ \hline \hline
$T$ [MeV] & $T=165.6\pm 4.5$ & $T=167.6\pm 4.6$ & \\ \hline
$\mu_B$ [MeV] & $\mu_B=28.5\pm 3.7$ & $\mu_B=28.9\pm 3.8$ & \\ \hline \hline
$\eta/\pi^-$ & $0.120 \pm 0.001$  & $0.112 \pm 0.001$ &   \\ \hline
$\rho^0/\pi^-$ & $0.114\pm 0.002$ & $0.135\pm 0.001$ & 
$0.183 \pm 0.028$  \cite{patricia2} (40-80\%)   \\ \hline
$\omega/\pi^-$ & $0.108 \pm 0.002$  & $0.102 \pm 0.002$ &   \\ \hline
$K^*(892)/\pi^-$ & $0.057\pm 0.002$ & $0.054 \pm 0.002$ &    \\ \hline
$\phi/\pi^-$ & $0.025 \pm 0.001$  & $0.024 \pm 0.001$ &   \\ \hline
$\eta^\prime/\pi^-$ & $0.0121 \pm 0.0004$  & $0.0115 \pm 0.0003$ &   \\ \hline
$f_0(980)/\pi^-$ & $0.0102 \pm 0.0003$ & $0.0097\pm 0.0003$
& $0.042 \pm 0.021$ \cite{patricia2} (40-80\%)   \\ \hline
$K^*(892)/K^-$ & $0.33\pm 0.01$ & $0.33\pm 0.01$
& 
\begin{tabular}{c} 
$0.205 \pm 0.033$ \cite{patricia2} (0-10\%)      \\
$0.219 \pm 0.040$ \cite{patricia2} (10-30\%)     \\
$0.255 \pm 0.046$ \cite{patricia2} (30-50\%)     \\
$0.269 \pm 0.047$ \cite{patricia2} (50-80\%)     \\
\end{tabular} \\ \hline
$\phi/K^*(892)$ & $0.446\pm 0.003$ & $0.448\pm 0.002$
& 
\begin{tabular}{c} 
$0.595 \pm 0.123$ \cite{patricia2} (0-10\%)      \\
$0.633 \pm 0.138$ \cite{patricia2} (10-30\%)     \\
$0.584 \pm 0.132$ \cite{patricia2} (30-50\%)     \\
$0.528 \pm 0.106$ \cite{patricia2} (50-80\%)     \\
\end{tabular}  \\ \hline
$\Lambda(1520)/\Lambda$ & $0.061\pm 0.002$ & $0.062\pm 0.002$
& 
\begin{tabular}{c} 
$0.022 \pm 0.010$ \cite{ludovic} (0-7\%)       \\
$0.025 \pm 0.021$ \cite{ludovic} (40-60\%)     \\
$0.062 \pm 0.027$ \cite{ludovic} (60-80\%)     \\
\end{tabular} \\ \hline
$\Sigma(1385)/\Sigma$ & $0.484\pm 0.004$ & $0.485\pm 0.004$
& $$   \\ \hline
\end{tabular}
\caption{Predictions of the thermal model for the ratios of hadronic resonances
  measured at RHIC at $\sqrt{s_{NN}}= 200$ GeV. In the two model
  calculations, the two different effective masses of the $\rho$ meson
  have been assumed. In the first case (second column) $m^*_\rho =
  770$ MeV, whereas in the second case (third column) $m^*_\rho = 700$
  MeV.  The two thermodynamic parameters have been fitted only to the
  ratios of stable hadrons. In the first case they are the same as
  those discussed in Sec. 2.  The preliminary experimental data are
  taken from \cite{patricia2,ludovic}.  Both the experimental and the
  theoretical pion yields are corrected for the weak decays. }
\end{centering}
\label{tab:fit700}
\end{table*}

\section{Ratios of the resonance yields \label{sec:ratios}}

In this and the next Section we present the results of the
single-freeze-out model for the ratios of the resonance yields and the
resonance transverse-momentum spectra. In our analysis we shall
consider two cases: in the first (standard) case we assume that the
masses of hadrons are not changed by the in-medium effects, whereas in
the second case we assume that the mass of the $\rho$ meson at
freeze-out is smaller than its vacuum mass. Inspired by the
preliminary STAR results \cite{patricia1,patricia2}, we shall use for
this purpose a rather low value of $m^*_\rho = 700$~MeV.  In these two
cases the values of the thermodynamic parameters are determined from
the experimental ratios of the yields of stable hadrons (those listed
in Table 1 of Ref. \cite{abwbwf}). This gives $T=165.6\pm 4.5$ MeV and
$\mu_B=28.5\pm 3.7$ MeV for the standard case \cite{abwbwf}, and
$T=167.6\pm 4.6$ MeV and $\mu_B=28.9\pm 3.8$ MeV for the case with the
modified mass of the $\rho$ meson. The reason for not including the
ratios of the resonance yields as the input in our appraoch is
twofold: firstly, the data describing the production of the resonances
are preliminary; secondly, there are no measurements of the
resonance spectra in the central events. For example, the data on
$\rho^0$ production were collected for the centrality class 40-80\%.  


We take into consideration the modification of the mass of the $\rho$
meson only, since at the moment there are no experimental hints
concerning the behaviour of the masses of other resonances (except for
the mass of $K^*(892)^0$ which is not changed). In Ref.
\cite{mmwfwb-inmedium} we studied the effect of the in-medium mass and
width modifications on the outcome of the thermal analysis in the
situation where a common scaling of baryon and meson masses with the
temperature or the density was assumed (only the masses of the
pseudo-Goldstone bosons were kept constant). The results of Ref.
\cite{mmwfwb-inmedium} showed that moderate modification 
particle masses are admissible. A satisfactory description of the
ratios of hadron abundances measured at CERN SPS may be obtained in a
thermal approach with the modified masses of hadronic resonances,
however, the changes of the masses affect the values of the optimum
thermodynamic parameters.  Similar results were obtained from the
analysis of the first RHIC data \cite{mm}. The effects of the
in-medium mass and width modifications on the outcome of the thermal
analysis were also studied in Refs.
\cite{wfwb-inmedium,hirschegg-inmedium,zschiesche1,zschiesche2,renk}.

Our predicitions for the ratios of the resonance yields are presented
in Table 1. The most interesting are the results for the
$\rho^0/\pi^-$ ratio which is 0.11 for the case without the in-medium
modifications and 0.14 for the case with the modified rho mass. The
first value is in a good agreement with a theoretical number presented
recently by Rapp \cite{Rapp}. We can see that the effect of the
dropping $\rho$-meson mass helps us to get closer to the preliminary
experimental result of 0.18, however the theoretical and the
experimental values still differ by more than one standard deviation.
Our model value for the ratio $f_0(982)/\pi^-$ is a factor of four
smaller than central value of the preliminary experimental result. On
the other hand, our results for $K^*(892)/K^-$ and
$\Lambda(1520)/\Lambda$ are larger than the preliminary experimental
values, whereas the $\phi/K^*(892)$ ratio agrees rather well with the
experiment ({\em cf.} Table~1).

A comparison of the theoretical and experimental results for the
ratios involving resonances is interesting from the point of view of
the discussion on the decoupling temperature, $T_{\rm therm}$, {\em
  i.e.}, the temperature characterizing the thermal freeze-out. If
$T_{\rm therm}$ is significantly smaller than $T_{\rm chem}$ one
expects \cite{rt1,rt2,heinzqgp} that the measured ratios such as
$K^*/K$ or $\rho^0/\pi^-$ are smaller than the values determined at
the chemical freeze-out. This behavior is due to the readjustment of
the resonance abundances (formed in elastic collisions) to the
decreasing temperature on the path the system follows from $T_{\rm
  chem}$ down to $T_{\rm therm}$. Our results shown in Table 1
indicate that certain experimental ratios are indeed smaller (such as
$K^*(892)/K^-$ or $\Lambda(1520)/\Lambda$ for central collisions),
whereas some are larger ($\rho^0/\pi^-$ and $f_0(982)/\pi^-$).
Consequently, from the comparison of the theoretical and experimental
ratios, at the moment, it is hard to draw a definite conclusion about
the difference between the two freeze-outs. The study gets even more
involved in view of possible in-medium modifications of the masses,
which affect the ratios and may complicate the overall thermodynamic
picture.

\begin{figure}[tb]
\begin{center}
\includegraphics[width=7.5cm]{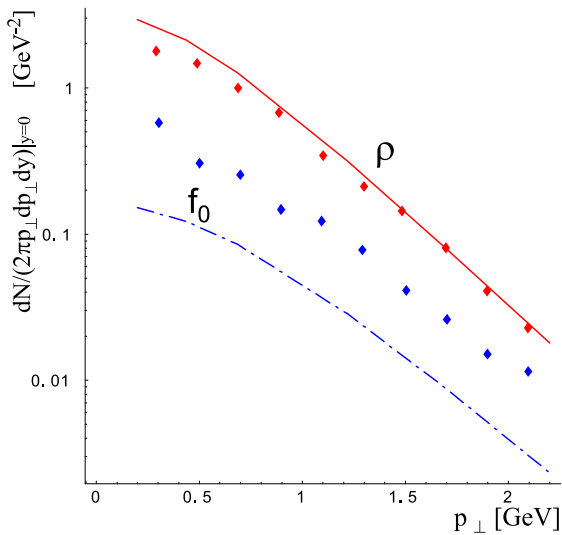}
\end{center}
\caption{The mid-rapidity 
  transverse-momentum spectra of the $\rho^0$ and $f_0(980)$ mesons,
  as obtained from the single-freeze-out model. The model parameters,
  $\tau=5$~fm and $\rho_{\rm max}=4.2$~fm, correspond to centralities
  $40-80$~\% \cite{abwbwf}. The data points are taken from
  Ref.~\cite{patricia2}. The vacuum value of the $\pi \pi$ phase shift
  is used in the model calculation, with the scaled phase shift
  yielding very similar results.}
\label{fig:rhof0}

\end{figure}

\section{Transverse-momentum spectra \label{sec:trspectra}}

In this Section we present the single-freeze-out model results for the
transverse-momentum spectra of various hadronic resonances. The method
of the calculation is the same as that used in the calculation of the
spectra of stable hadrons \cite{wbwf,str,zakop}. In particular, the
feeding of the resonance states from all known higher excited states is
included, which leads to the enhancement of the resonance production
characterized by the factors $d_i$. As usual, the two thermodynamic
parameters and the two geometric parameters (fitted separately for
different centrality windows) are taken from Ref.  \cite{abwbwf}.
Knowing the centrality dependence of $\tau$ and $\rho_{\rm max}$ we
may analyze the resonance production at different centralities and
compare it to the existing data (note that the data on $\rho$
production are collected only for rather peripheral collisions
corresponding to the centrality window 40-80\%).

\begin{figure}[tb]
\begin{center}
\includegraphics[width=8.7cm]{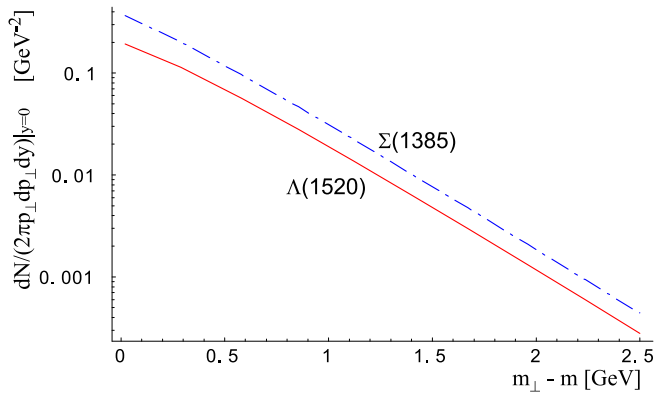}
\end{center}
\caption{The single-freeze-out model 
  predictions for the mid-rapidity transverse-mass spectra of
  $\Lambda(1520)$ and $\Sigma(1385)$.  The model parameters,
  $\tau=8.7$~fm and $\rho_{\rm max}=7.2$~fm, correspond to the most
  central events, $c=0-20$\% \cite{abwbwf}.  }
\label{fig:ls}
\end{figure}

\begin{figure}[tb]
\begin{center}
\includegraphics[width=8.5cm]{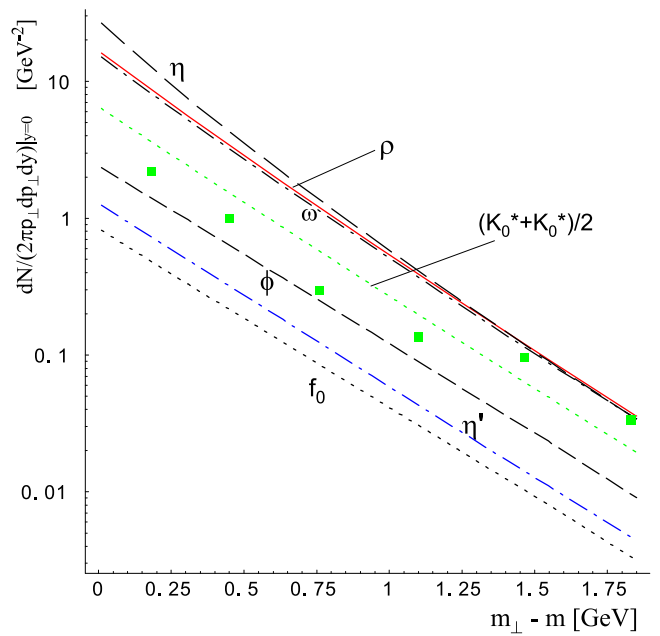}
\end{center}
\caption{The mid-rapidity 
  transverse-momentum spectra of the various resonances, as obtained
  from the single-freeze-out model. The preliminary data for the $K^\ast$ 
are taken from Ref.~\cite{starKstar1}. The model parameters, $\tau=8.7$~fm
  and $\rho_{\rm max}=7.2$~fm, correspond to centralities $0-20$~\%
  \cite{abwbwf}.}
\label{fig:res}
\end{figure}

In Fig. \ref{fig:rhof0} we show our results for the
transverse-momentum spectra of $\rho^0$ and $f_0(892)$, and compare
them to the preliminary data obtained by the STAR Collaboration
\cite{patricia2}. The expansion parameters are $\tau=5$~fm and
$\rho_{\rm max}=4.2$~fm, which corresponds to the centrality of 40-80\%
\cite{abwbwf}. In the presented calculation the vacuum value of the
$\pi \pi$ phase shift has been used.  We have checked that the scaled phase
shift gives very similar results, and the small change of the
thermodynamic parameters due to the drop of the rho mass
affects the spectrum very little. The results shown in Fig.~\ref{fig:rhof0} 
indicate that our model can quite well reproduce the
experimental spectrum of the rho meson and the slope of the $f_0(980)$
spectrum. We find, however, the discrepancy in the normalization 
between the model and the preliminary 
experimental data for the $f_0(980)$ spectrum, which reflects
the result of Table 1. In
Fig. ~\ref{fig:ls} we show our predictions for the spectra of
$\Lambda(1520)$ and $\Sigma(1350)$. Since $\Lambda(1520)$ is currently
measured in the central collisions \cite{ludovic}, in this case we
used the values of the geometric parameters corresponding to the most
central collisions, 
namely $\tau=8.7$~fm and $\rho_{\rm max}=7.2$~fm, which corresponds to centrality of 
$0-20$\% \cite{abwbwf}. In Fig. \ref{fig:ls} we also show
the spectrum of $\Sigma(1385)$ which might be measured by STAR in the
near future \cite{christelle}.  In Fig.~\ref{fig:res} we compare 
our predictions for the spectrum of $K^\ast(892)$ to the preliminary 
experimental results \cite{starKstar1}. To complete our discussion, we also
show our predictions for other resonances for the most-central case, 
namely $\eta$, $\rho$, $\omega$, $\phi$,
and $\eta^\prime$ and $f_0$.

\section{Conclusion \label{sec:conclude}}

In this paper we have presented a comprehensive analysis of resonance 
production at RHIC in the thermal approach. We have studied the $\pi^+ \pi^-$
invariant mass spectra, the ratios of the resonances, and their transverse-momentum
spectra. 

Our calculation of the invariant mass
distribution of the $\pi^+ \pi^-$ pairs, performed in the framework of the 
single-freeze-out model, has been done with the use
of the derivatives of the experimental phase shifts (and not the spectral
function), which guarantees
the thermodynamic consistency of our approach. Decays of higher 
resonances have been included in a complete fashion. Moreover, we have taken
carefully into account the realistic kinematic cuts corresponding to
the STAR experimental acceptance.  We find that the
preliminary STAR result indicating a drop of the rho meson mass down
to 700 MeV cannot be reproduced in our model when the vacuum 
properties of the $\rho$ are used.
A good qualitative agreement is found, however, when 
the mass of the $\rho$ is lowered. Therefore our 
calculation confirms the {\em dropping-mass scenario} for the 
$\rho$, induced by the medium effects.

We stress that the shape of the $\pi^+ \pi^-$ spectrum is very
sensitive to the freeze-out temperature.  In a sense, it can be used
as a ``termometer'', independent of other methods of determining the
temperature, such as the study of the ratios of the stable particles,
or the slopes of the transverse momenta.  We have shown that at small
$T$ the contributions from the low lying resonances are enhanced, and
the contributions at high mass are suppressed.  On the contrary, at
large $T$ the contributions from the high lying states are enhanced
and become visible (like, {\em e.g.}, the contribution from
$f_2(1270)$), while the low mass contributions are reduced.  In the
single-freeze-out model, although the freeze-out process takes place
at relatively high temperature $T \sim$ 165 MeV, the decays of the
resonances lead to an effective ``cooling'' of the spectrum, with
low-mass resonances acquiring more feeding from the higher states than
the high-mass resonances.  This would be equivalent to taking a lower
temperature in a naive calculation without the resonance decays. Such
effects, seen initially in the shape of the transverse-momentum
spectra, are therefore also seen in the invariant mass $\pi^+ \pi^-$
distributions.

The experimental information on the resonance production is crucial
for our understanding of the freeze-out process. In particular, it may
be used to distinguish between the scenario with the two freeze-outs,
separate for the elastic and inelastic processes, and the scenario
with a single freeze-out.
For example, in the scenario with the two distinct freeze-outs the ratios
involving the resonace yield in the numerator and the stable-hadron
yield in the denominator ({\em e.g.}, $\rho/\pi$, $K^*/K$, or
$\Lambda(1520)/\Lambda$) should be smaller than the analogous ratio calculated at
the chemical freeze-out. By comparing the preliminary STAR data to the
predictions of the thermal model we have pointed out that certain ratios of
this type are indeed somewhat smaller, while other are larger.  Consequently, at
the present time one cannot exclude which approximation, the one with two or the one with 
one freeze-out, is more appropriate in the thermal approach. More acurate data would be 
highly desired in that regard.

\begin{acknowledgements}
  We are grateful to Patricia Fachini for valuable discussions and
  technical information on the kinematic cuts in the STAR experiment.
We are also grateful to Christell Roy and Ludovic Gaudichet
for explanations concering the $\Lambda(1520)$ production.
 WB acknowledges the support of
  Fundacao para Ciencia e Technologia, grant PRAXIS
  XXI\-/BCC\-/429\-/94.  WB's and WF's research was partially
  supported by the Polish State Committee for Scientific Research
  grant 2~P03B~09419. BH acknowledges the support of Fundacao para
  Ciencia e Technologia, grant POCTI/35304/2000.
\end{acknowledgements}

\appendix 

\section{Parameters for the $\pi\pi$ phase shifts \label{app:param}}

The following parameterization is used for the $\pi\pi$ phase shifts
of Eq.~(\ref{tandel}) \cite{pipipar} (all in units of
$m_{\pi^+}=139.57$~MeV):
\begin{eqnarray} 
A_0^0 &=& 0.220, \; A^1_1=0.379\cdot 10^{-1}, \; A^2_0= -0.444\cdot 10^{-1}, 
\nonumber \\
B_0^0 &=& 0.268, \; B^1_1=0.140\cdot 10^{-4}, \; B^2_0= -0.857\cdot 10^{-1}, 
\nonumber \\
C_0^0 &=&-0.0139, \; C^1_1= -0.673\cdot 10^{-4}, \; C^2_0= -0.00221, \nonumber \\
D_0^0 &=&-0.00139,  D^1_1= 0.163\cdot 10^{-7},  D^2_0= -0.129\cdot 10^{-3}, 
\nonumber \\
s_0^0 &=& 36.77, \; s^1_1=  30.72, \; s^2_0= -21.62. \label{parcol}
\end{eqnarray}

\section{Kinematic cuts for two-body decays \label{app:two}}

Let the unlabeled quantities refer to the decaying resonance, and
labels $1$ and $2$ to the decay products.  The number of pairs 
coming from the decays of the resonance formed on the
freeze-out hypersurface is \cite{ornik,zakop}
\begin{equation}
{N}_{12}=\int \frac{d^{3}p_{1}}{E_{1}}C_{1}\int \frac{%
d^{3}p}{E}C C_{2}B\left( p, p_{1}\right) S(p),
\label{npk}
\end{equation}
where the symbols $C$, $C_{1}$, and $C_2$ denote kinematic cuts, and
the source function of the decaying resonance is obtained from the
Cooper-Frye formula \cite{CF1,CF2},
\begin{equation}
S(p) = \int d\Sigma _{\mu }\left( x\right) \,p^{\mu
}\,\,f \left[ p\cdot u\left( x\right) \right] .\label{npk2}
\end{equation}
The quantity $f$ is the thermal distribution of the particle decaying
at the hypersurface $\Sigma$, with the collective flow described by
the four-velocity $u$. We pass to rapidity and transverse momentum
variables in the laboratory frame, and have explicitly
\begin{eqnarray}
&&B(p ,p_{1}) ={\frac{b}{4\pi p_{1}^{\ast }}}\delta \left( {%
\frac{p \cdot p_{1}}{m}}-E_{1}^{\ast }\right) 
\label{B} \\
&&={\frac{b}{4\pi p_{1}^{\ast }}}\delta \left( {\frac{m^{\perp }m_{1}^{\perp }
\cosh (y-y_{1})-p^{\perp }p_{1}^{\perp }\cos \gamma }{m}}-E_{1}^{\ast
}\right) ,  \nonumber
\end{eqnarray}
where $p_{1}^{\ast }$ and $E_{1}^{\ast }$ denote the momentum and
energy of particle $1$ in the rest frame of the decaying resonance,
$\gamma$ is the angle between ${\bm p}^\perp$ and ${\bm p}_1^{\perp}$
in the laboratory frame, and $b$ is the branching ratio for the
considered decay channel.

In general, for the considered cylindrical symmetry, we have 
\begin{eqnarray}
C &=&\theta (p_{\rm low}^{\perp }\leq p^{\perp }\leq p_{\rm
high}^{\perp })\theta (y_{\rm low}\leq y \leq y_{\rm high}),  \label{cuts}
\\
C_{i} &=&\theta (p_{i ,{\rm low}}^{\perp }\leq p_{i}^{\perp
}\leq p_{i, {\rm high}}^{\perp })\theta (y_{i, {\rm low}}\leq y \leq y_{i, {\rm high}}), 
\nonumber
\end{eqnarray}
with $i=1,2$ and $\theta (a\leq x\leq b)=1$ if the condition is satisfied, and $0$
otherwise. Due to momentum conservation 
we should use in (\ref{cuts})
\begin{eqnarray}
p^{\perp}_2&=&\sqrt{\left( p^{\perp }\right) ^{2}+\left( p_{1}^{\perp }\right)
^{2}-2p^{\perp }p_{1}^{\perp }\cos \gamma } \nonumber \\
y_2 &=&\sinh ^{-1}\left [ \frac{m^{\perp
}\sinh y-m_{1}^{\perp }\sinh y_{1}}{m_\perp} \right ]. \label{momcon}
\end{eqnarray}

Next, we perform the integration over $\gamma$ in Eq.~(\ref{npk},\ref{B}). 
The $\delta$ function gives the condition $\cos \gamma =\cos
\gamma _{0}$, with 
\begin{equation}
\cos \gamma _{0}=\frac{m^{\perp }m_{1}^{\perp }\cosh
(y-y_{1})-m E_{1}^{\ast }}{p^{\perp}p_{1}^{\perp }},
\end{equation}
which takes effect only if $-1\leq \cos \gamma _{0}\leq 1$. A factor of $2$
follows from the two solutions of $\gamma =\arccos (\cos \gamma _{0})$ for $%
\gamma \in \lbrack 0,2\pi )$. The final result is 
 \begin{eqnarray}
{N}_{12} &=&{\frac{b m}{p_{1}^{\ast }}}
\int_{p_{1, \rm{low}}^{\perp }}^{p_{1, {\rm high}}^{\perp }}
\!\! dp_{1}^{\perp }\int_{y_{1, {\rm low}}}^{y_{1, {\rm high}}} \!\! dy_{1}
\int_{p_{\rm low}^{\perp }}^{p_{\rm high}^{\perp }} \!\! dp^{\perp } 
\int_{y_{\rm low}}^{y_{\rm high}} 
\!\! dy  \nonumber \\
&&\times C_{2}^{0}\,\frac{\theta (-1\leq \cos \gamma _{0}\leq 1)}
{\sin \gamma _{0}} S(p^{\perp }),
\label{fin2}
\end{eqnarray}
where we have introduced 
\begin{eqnarray}
C_{2}^{0} =\left. C_{2}\right| _{\cos \gamma =\cos \gamma _{0}},
\end{eqnarray}
with the substitution (\ref{momcon}) understood.

The experimental cuts frequently involve cuts on pseudo-rapidity of particles. 
This amount to adding extra conditions of the form 
\begin{eqnarray}
&& \eta_i^{\rm low} \leq {1 \over 2} \log\left( 
\frac{\sqrt{p_i^{\perp 2}+m_i^{\perp 2}{\rm sinh}^2 y_i}+m_i^{\perp 2}{\rm sinh}^2 
y_i}
{\sqrt{p_i^{\perp 2}+m_i^{\perp 2}{\rm sinh}^2 y_i}-m_i^{\perp 2}{\rm sinh}^2 y_i}
\right) 
\nonumber \\ &&\;\;\;\;\; \leq \eta_i^{\rm high},
\end{eqnarray}
which may be implemented in Eq.~(\ref{fin2}) in the form of $\theta$
functions, denoted by $C^\eta_i$.

The above formulas have been written for a sharp resonance. For a wide resonance, 
according to Eq.~(\ref{master}), the generalization assumes the form (\ref{gener}).

\section{Kinematic cuts for three-body decays \label{app:three}}

The three-body decay is usually considered in the rest frame of the
resonance of mass $m$. Here, since the cuts are defined in the
laboratory frame, we need to consider the kinematics in this frame.
The phase-space integral for the decay of particle of mass $m$ and
momentum ${\bm p}$ into products $1$, $2$, and $3$ is proportional to
\begin{eqnarray}
&&\int \frac{d^{3}p_1}{E_1}\frac{d^{3}p_2}{E_2} \frac{d^{3}p_3}{E_3} C \delta 
\left( E - E_1 - E_2 - E_3\right) \nonumber \\
&&\times \delta ^{(3)}\left( {\bm p}-
{\bm p}_{1} - {\bm p}_{2} - {\bm p}_{3}\right) \left| \cal{M}\right| ^{2},  
\label{phsp}
\end{eqnarray}
where we have introduced $E=\sqrt{m^2+{\bm p}^2}$, {\em etc}, and $C$
denotes a generic cut in the kinematic variables, to be specified
later.  For simplicity we assume that $\cal{M}$ can be approximated by
a constant, \emph{i.e.} only the phase-space effect is included. This
condition can be relaxed at no difficulty, if needed.  We are
interested in the invariant-mass distribution of particles $1$ and
$2$, hence we introduce the $\delta[M-\sqrt{(E_1+E_2)^2-({\bm
    p}_1+{\bm p}_2)^2}]$ function in Eq.~(\ref{phsp}) and obtain the
expression for the probability of emitting a pair of invariant mass
$M$ from a particle moving with momentum ${\bm p}$:
\begin{eqnarray}
\frac{dN_{12}({\bm p})}{dM} &=& a \int \frac{d^{3}p_{1}}{E_{{1}}}\frac{d^{3}p_{2}}{E_{{2}}}\frac{%
d^{3}p_{3}}{E_{{3}}}C \delta \left( E - E_{{1}} - E_{{2}} - E_{{3}}\right) \nonumber \\
& \times &\delta \left ( M-\sqrt{(E_1+E_2)^2-({\bm p}_1+{\bm p}_2)^2} \right ) \nonumber \\
& \times & \delta ^{(3)}\left( \mathbf{p}-
\mathbf{p}_{1} - \mathbf{p}_{2} - \mathbf{p}_{3}\right),  \label{phsp2}
\end{eqnarray}
where $a$ is a normalization constant. First, we trivially carry the
integration over ${\bm p}_3$ through the use of the last $\delta$ in
Eq.~(\ref{phsp2}). Next, we pass to the rapidity and
transverse-momentum variables, and carry the integration over the
angle between momenta ${\bm p}^\perp_1$ and ${\bm p}^\perp_2$, denoted
as $\gamma$. This yields
\begin{eqnarray}
\frac{dN_{12}({\bm p}^\perp, y) }{dM} &=& 
2\pi a \sum_\epsilon \int p_1^\perp dp_1^\perp dy_1d \alpha   
\int p_2^\perp dp_2^\perp dy_2 C
\nonumber \\ & \times& \frac{1}{A} \frac{M}{p_1^\perp p_2^\perp |\sin \gamma_0|} 
\delta \left( A - E_{{3}}\right), 
 \label{phsp3}
\end{eqnarray}
where $\alpha$ is the angle between ${\bm p}^\perp$ and ${\bm p}^\perp_1$,
\begin{eqnarray}
\cos \gamma_0 = 
\frac{-M^2+m_1^2+m_2^2+2m_1^\perp m_2^\perp {\rm cosh}(y_1-y_2)}{2 p_1^\perp p_2^\perp}, \nonumber \\
\label{gamma0}
\end{eqnarray}
and the sum over $\epsilon$ results from the two solutions for $\gamma_0$, 
{\em i.e.} $\sin \gamma_0=\epsilon \sqrt{1+\cos^2 \gamma_0}$. Next, 
\begin{eqnarray}
A=m^\perp {\rm cosh} y  - m_1^\perp {\rm cosh} y_1  
 - m_2^\perp {\rm cosh} y_2, \label{Adef}
\end{eqnarray} 
and, finally,
\begin{eqnarray}
E_3 &=&\left [ B
-2 p^\perp p^\perp_1 \cos \alpha \right . \label{E3} \\
&& \left . -2 p^\perp p^\perp_2 \cos (\alpha-\gamma_0)
+2 p_1^\perp p^\perp_2 \cos \gamma_0 
 \right ]^{1/2},  \nonumber 
\end{eqnarray}
with
\begin{eqnarray}
B &=& m_3^2+(m^\perp {\rm sinh}y-m^\perp_1 
{\rm sinh}y_1-m^\perp_2 {\rm sinh}y_2)^2 \nonumber \\
&& + p^{\perp 2}+  p_1^{\perp 2}+  p_1^{\perp 2}.
\end{eqnarray}
We need still to carry the integration over the angle $\alpha$.  We
square the expression under the $\delta$ function in
Eq.~(\ref{phsp3}), obtaining
\begin{eqnarray}
&& A^2 = B -2 p^\perp p^\perp_1 \cos \alpha \\
&& -2 p^\perp p^\perp_2 ( \cos \alpha \cos \gamma_0 -\sin \alpha \sin \gamma_0)
+2 p_1^\perp p^\perp_2 \cos \gamma_0 .\nonumber
\end{eqnarray}
The squaring imposes the condition 
\begin{equation}
A \ge 0 .
\label{condA}
\end{equation}
This equation 
can be solved straightforwardly by introducing $t=\tan(\alpha/2)$,
\begin{eqnarray}
\cos \alpha &=&\frac{1-t^2}{1+t^2},\nonumber \\ 
\sin \alpha &=&\frac{2t}{1+t^2}, \label{half}
\end{eqnarray}
and the notation
\begin{eqnarray}
G&=&\frac{-A^2+B+2 \cos \gamma_0 p_1^\perp p_2^\perp}{p^\perp p^\perp_2},\nonumber \\
H&=& \cos \gamma_0+\frac{p^\perp_1}{p^\perp_2} . \label{CD}
\end{eqnarray}
Now Eq.~(\ref{E3}) acquires the simple quadratic form
\begin{eqnarray}
(1+t^2)G=2t \epsilon \sqrt{1-\cos^2 \gamma_0} +(1-t^2)H, 
\end{eqnarray}
with the solutions 
\begin{eqnarray}
t_0(\epsilon,\epsilon')&=&\tan(\alpha_0(\epsilon,\epsilon')/2) \label {a0}\\
&=&\frac{\epsilon \sqrt{1-\cos^2\gamma_0}+\epsilon' 
\sqrt{H^2-G^2+\sin^2\gamma_0}}{G+H}, \nonumber
\end{eqnarray}
with $\epsilon'=\pm 1$.
The solutions make sense under the condition 
\begin{equation} 
H^2-G^2+\sin^2\gamma_0 \ge 0.
\label{condCD}
\end{equation}
From the derivative of the delta function we obtain the factor
\begin{eqnarray}
\frac{A}{p^\perp p_2^\perp |H \sin \alpha_0(\epsilon,\epsilon')-
\epsilon \sin \gamma_0 \cos\alpha_0(\epsilon,\epsilon')|}.
\end{eqnarray} 
It is easy to check that this factor is independent of $\epsilon$ and
$\epsilon'$.  Thus, to the extent that the cut function $C$ does not
involve azimuthal angles, we may use one combination of these signs
and put a factor of four. The final result is
\begin{eqnarray}
\frac{dN_{12}({p}^\perp, y) }{dM} &=& 8\pi a   \int dp_1^\perp dy_1 
\int dp_2^\perp dy_2 C
\nonumber \\ 
& \times & \theta(A) \theta(H^2-G^2+\sin\gamma_0^2 ) \label{phsp4}\\
& \times& \frac{M}{|\sin \gamma_0|} 
\frac{1}{p^\perp p_2^\perp |H \sin \alpha_0 -\sin \gamma_0 \cos\alpha_0|}, 
\nonumber 
\end{eqnarray}  
where $\alpha_0$ is any of the angles (\ref{a0}), and 
all the necessary substitutions are understood. 
The normalization constant $a$ can be obtained from the condition 
\begin{eqnarray}
\int dM \frac{dN_{12}({p}^\perp, y)}{dM} =1,
\end{eqnarray}
with no cuts present, {\em i.e.} with the cut function set to unity,
$C=1$.

The form of the cut function $C$ involves the ranges in the
integration variables $p_1^\perp$, $y_1$, $p_2^\perp$, $y_2$, the cuts
on the pseudo-rapidity of particles $1$ and $2$, as well as cuts on the
rapidity and the transverse momentum of the pair of particles $1$ and
$2$ \cite{FachPriv}. These cuts assume a simple form of products of the
$\theta$ functions. Then, Monte Carlo methods are appropriate to
compute Eq.~(\ref{phsp4}).


\begin{thebibliography}{99}
  
\bibitem{starKstar1} P. Fachini, STAR Collaboration, J. Phys. G {\bf
    28}, 1599 (2002).
  
\bibitem{starKstar2} C. Adler et al., STAR Collaboration, Phys. Rev. C
  {\bf 66}, 061901 (2002).
  
\bibitem{patricia1} P.~Fachini, STAR Collaboration, Nucl. Phys. A {\bf
    715}, 462c (2003).

\bibitem{patricia2} P.~Fachini, STAR Collaboration, nucl-ex/0305034
 and private communication.
  
\bibitem{christelle} C.~Roy, nucl-ex/0303004.
  
\bibitem{markert} C. Markert, Proceedings of the 19th Winter Workshop
  on Nuclear Dynamics, Breckenridge, Colorado (USA) (2003).

  
\bibitem{ludovic} L.~Gaudichet, Proceedings of the 7th International
  on Strangeness in Quark Matter, North California (USA) (2003), and
  private communication.

  
\bibitem{BR} G. Brown and M. Rho, Phys. Rev. Lett. {\bf 66}, 2720
  (1991).
  
\bibitem{hatlee} T. Hatsuda and S.~H. Lee, Phys. Rev. C {\bf 46}, R34
  (1993).
  
\bibitem{CERES} CERES Collaboration, G. Agakichiev et al., Phys. Rev.
  Lett.  {\bf 75} (1995) 1272.
  
\bibitem{HELIOS} HELIOS/3 Collaboration, M. Masera et al., Nucl. Phys.
  A {\bf A590} (1995) 93c.
  
\bibitem{shuryakbrown} E.~V.~Shuryak and G.~E.~Brown, Nucl. Phys. A
  {\bf 717}, 322 (2003).

\bibitem{KolbP} P. F. Kolb and M. Prakash, Phys. Rev. C {\bf 67}, 044902 (2003). 


\bibitem{Rapp} R. Rapp,  hep-ph/0305011.


\bibitem{wbwf} W. Broniowski and W. Florkowski, Phys. Rev. Lett. {\bf
    87}, 272302 (2001).

\bibitem{str} W. Broniowski and W. Florkowski, Phys. Rev. C {\bf 65},
  064905 (2002).

\bibitem{zakop} W. Broniowski, A. Baran, and W. Florkowski, Acta.
  Phys. Pol. B {\bf 33}, 4235 (2002).
  
\bibitem{Da1} R.~Dashen, S.~Ma, and H.~J.~Bernstein, Phys. Rev. {\bf
    187}, 345 (1969).
  
\bibitem{Da2} R.~F.~Dashen and R.~Rajaraman, Phys. Rev. D {\bf 10},
  694 (1974); R.~F.~Dashen and R.~Rajaraman, Phys. Rev. D {\bf 10},
  708 (1974).
  
\bibitem{Wmsc} W. Weinhold, {\it Zur Thermodynamik des
    Pion-Nukleon-Systems}, Diplomarbeit, TH Darmstadt, Sept. 1995.
  
\bibitem{WFNapp} W. Weinhold, B.~L.~Friman, and W. N\"orenberg, Acta
  Phys. Pol. B {\bf 27}, 3249 (1996).

\bibitem{WFNnote} W. Weinhold, B.~L.~Friman, and W. N\"orenberg,
     {\it Thermodynamics with resonance states},
     GSI report 96-1, p. 67. 
  
\bibitem{WFNplb} W. Weinhold, B.~L.~Friman, and W. N\"orenberg,
     Phys.  Lett. B {\bf 433}, 236 (1998).
  
\bibitem{Wphd} W. Weinhold, {\it Thermodynamik mit
    Resonanzzust\"anden}, Dissertation, TU Darmstadt, 1998.

\bibitem{denis} K.~G.~Denisenko and St. Mr\'owczy\'nski, Phys. Rev. C {\bf 35},
1932 (1987).

\bibitem{larionov} A.~B.~Larionov, W.~Cassing, M.~Effenberger, and U.~Mosel, 
Eur. Phys. J. A {\bf 7}, 507 (2000).

\bibitem{pelaez} J.~R.~Pel\'aez, Phys. Rev. D {\bf 66}, 096007 (2002). 
  
\bibitem{pipipar} G.~Colangelo, J.~Gasser, and H.~Leutwyler, Nucl.
  Phys. B {\bf 603}, 125 (2001).
  
\bibitem{wfwbmm} W. Florkowski, W. Broniowski, and M. Michalec, Acta
  Phys. Pol. B {\bf 33}, 761 (2002).

\bibitem{BM4} P. Braun-Munzinger, D. Magestro, K. Redlich, and J.
  Stachel, Phys. Lett. B {\bf 518}, 41 (2001).

\bibitem{review} P. Braun-Munzinger, K. Redlich, and J. Stachel,
  nucl-th/0304013.

\bibitem{FachPriv} P. Fachini, private communication.

\bibitem{bjorken} J. D. Bjorken, Phys. Rev. D {\bf 27}, 140 (1983).

\bibitem{baym} G. Baym, B. Friman, J.-P. Blaizot, M. Soyeur, and W.~Czy\.z,
Nucl. Phys. A {\bf 407}, 541 (1983). 

\bibitem{Kolya} P. Milyutin and N. N. Nikolaev, Heavy Ion Phys {\bf 8}, 
333 (1998); V. Fortov, P.~Milyutin, and N. N. Nikolaev,
JETP Lett. {\bf 68}, 191 (1998).

\bibitem{siemens} P. J. Siemens and J. Rasmussen, Phys. Rev. Lett. {\bf 42},
880 (1979); P. J. Siemens and J. I. Kapusta, Phys. Rev. Lett. {\bf 43},
1486 (1979).

\bibitem{SSH} E. Schnedermann, J. Sollfrank, and U. Heinz,
 Phys. Rev. C {\bf 48}, 2462 (1993).

\bibitem{BL} T. Cs\"{o}rg\H{o} and B. L\"{o}rstad, Phys. Rev. 
C {\bf 54}, 1390 (1996).

\bibitem{cs1} T. Cs\"{o}rg\H{o}, Heavy Ion Phys. {\bf 15}, 1 (2002).

\bibitem{Rischke} D. H. Rischke and M. Gyulassy, Nucl. Phys.
A  {\bf 697}, 701 (1996); Nucl. Phys. A {\bf 608}, 479 (1996).

\bibitem{SH} R. Scheibl and U. Heinz, Phys. Rev. C {\bf 59}, 1585
(1999).

\bibitem{cs2} T. Cs\"{o}rg\H{o}, F. Grassi, Y. Hama, and T. Kodama,
nucl-th/0305059.
  
\bibitem{brahmsy} D. Ouerdane, BRAHMS Collaboration, Nucl. Phys. A
  {\bf 715}, 478c (2003); I.~G.~Bearden et al., BRAHMS Collaboration,
  Phys. Rev.  Lett. {\bf 90}, 102301 (2003).

\bibitem{PDG} Particle Data Group, Eur. Phys. J. C {\bf 15} (2000) 1.

\bibitem{hagedorn} R. Hagedorn, Suppl. Nuovo Cim.  {\bf 3}, 147
  (1965); preprint CERN 71-12 (1971), preprint CERN-TH. 7190/94 (1994)
  and references therein.
  
\bibitem{myhag} W. Broniowski and W. Florkowski, Phys. Lett. B {\bf
    490}, 223 (2000).

\bibitem{bled} W. Broniowski, in Proc. of Few-Quark Problems, Bled,
  Slovenia, July 8-15, 2000, eds. B. Golli, M. Rosina, and S. \v
  Sirca, p. 14, hep-ph/0008112.
  
\bibitem{rafhag} A. Tounsi, J. Letessier, and J. Rafelski,
  contribution to the NATO Advanced Study Workshop on Hot Hadronic
  Matter: Theory and Experiment, Divonne-les-Bains, France, 27 Jun - 1
  Jul 1994, p. 105.
  
\bibitem{mmwfwb-inmedium} M. Michalec, W. Florkowski and W.
  Broniowski, Phys. Lett. B {\bf 520}, 213 (2001).

\bibitem{mm} M. Michalec, PhD Thesis, nucl-th/0112044.
  
\bibitem{abwbwf} A.~Baran, W.~Broniowski, and W.~Florkowski,
  nucl-th/0305075.

\bibitem{wfwb-inmedium} W. Florkowski and W. Broniowski, Phys. Lett. B
  {\bf 477}, 73 (2000).

\bibitem{hirschegg-inmedium} W. Florkowski and W. Broniowski,
  Proceedings of the International Workshop XXVIII on Gross Properties
  of Nuclei and Nuclear Excitations, Hirschegg, Austria, 2000, p. 275.
  
\bibitem{zschiesche1} D.~Zschiesche, L.~Gerland, S.~Schramm,
  J.~Schaffner-Bielich, H.~Stoecker, and W.~Greiner, Nucl. Phys. A
  {\bf 681}, 34 (2001).
  
\bibitem{zschiesche2} D.~Zschiesche, S.~Schramm, J.~Schaffner-Bielich,
  H.~Stocker, and W.~Greiner, Phys. Lett. B {\bf 547}, 7 (2002).

\bibitem{renk} T.~Renk, hep-ph/0210307.

  
\bibitem{rt1} G. Torrieri and J. Rafelski, J. Phys. G {\bf 28}, 1911
  (2002).

\bibitem{rt2} C. Markert, G. Torrieri, and J. Rafelski, {\it Campos do
    Jordao 2002, New states of matter in hadronic interactions,
    533}, hep-ph/0206260.

\bibitem{heinzqgp} U. Heinz, plenary talk at 16th International
  Conference on Particles and Nuclei (PANIC 02), Osaka, Japan, 30 Sep
  - 4 Oct 2002, nucl-th/0212004.
  
\bibitem{CF1} F. Cooper and G. Frye, Phys. Rev. D {\bf 10}, 186 (1974).
  
\bibitem{CF2} F. Cooper, G. Frye, and E. Schonberg, Phys. Rev. D {\bf
    11}, 192 (1975).

\bibitem{ornik} J. Bolz, U. Ornik, M. Pl\"umer, B.R. Schlei, and
R.M.~Weiner, Phys. Rev. D {\bf 47}, 3860 (1993).

\end{thebibliography}
\end{document}